# A LIGHT WEIGHT PROTOCOL TO PROVIDE LOCATION PRIVACY IN WIRELESS BODY AREA NETWORKS

Mohammed Mana[1], Mohammed Feham[1], and Boucif Amar Bensaber[2]

STIC Lab., Department of telecommunications, University of Tlemcen, Tlemcen, Algeria
`manamed_alg@yahoo.fr, m_feham@mail.univ-tlemcen.dz`

[2]Laboratoire de mathématiques et informatique appliquées LAMIA, Université du Québec à Trois-Rivières
C.P. 500 Trois-Rivières, Québec, Canada G9A 5H7
`Boucif.Amar.Bensaber@uqtr.ca`

## ABSTRACT

*Location privacy is one of the major security problems in a Wireless Body Area Networks (WBANs). An eavesdropper can keep track of the place and time devices are communicating. To make things even worse, the attacker does not have to be physically close to the communicating devices, he can use a device with a stronger antenna. The unique hardware address of a mobile device can often be linked to the identity of the user operating the device. This represents a violation of the user's privacy. The user should decide when his/her location is revealed and when not. In this paper, we first categorize the type of eavesdroppers for WBANs, and then we propose a new scheme to provide the location privacy in Wireless Body Area Networks (WBANs).*

## 1. INTRODUCTION

Location privacy has been always a prime concern in wireless sensor networks with regard to healthcare applications. Sending data out from a patient through wireless media can pose serious threats to the privacy of an individual [1].

Location privacy can be defined as the confidentiality of personal location information [2]. Location privacy is another kind of special privacy requirements due to the distinctiveness of location information, which can be obtained in many means (direct localization, calculation, or eavesdropping). Thus, traditional methods designed for data confidentiality cannot protect personal location privacy [3]. As far as the party is concerned, location privacy can be divided into two types: source (sender) location privacy or sink (recipient) location privacy.

Many schemes providing the anonymity of communication parties in Internet and Ad-hoc networks are not appropriate for wireless body area networks due to the nature of communicating devices which are very resource limited [4].

Also, the location privacy mechanisms employed in Wireless Sensor Networks do generally not offer the best solutions to be used in Wireless Body Area Networks for the latter have specific features that should be taken into account when designing the security architecture. The number of sensors on the human body, and the range between the different nodes, is typically quite limited. Furthermore, the sensors deployed in a WBAN are under surveillance of the person carrying these devices. This means that it is difficult for an attacker to physically access the nodes without this being detected. When designing location privacy protocols for WBAN, these characteristics should be taken into account in order to define optimized solutions with respect to the available resources in this specific environment [8].

Following are presented some schemes proposed in the literature to provide location privacy in this type of networks.

Gehrmann et al. [8] presented the Bluetooth anonymity mode. The authors propose to use three types of addresses: the fixed Bluetooth address, the active Bluetooth address and the alias address. Bluetooth devices working in anonymous mode use the active address for connection establishment and communication. It is a random 48-bit address that is changed regularly. The use of the fixed Bluetooth hardware address is still supported in the Bluetooth anonymity mode. This is done to allow direct connections between two trusted devices. However, the authors suggest combining page scanning based on the fixed Bluetooth hardware addresses with alias authentication. The Bluetooth anonymity mode does not provide full protection to location privacy attacks. Since the messages exchanged during a page scan contain the fixed Bluetooth hardware address and are not encrypted, a passive eavesdropper can easily detect that a particular device is present. Alias authentication is also not sufficient to avoid active tracking attacks. An adversary can perform a replay attack and force two devices to reuse old alias addresses. Since Bluetooth does not provide mechanisms to protect the integrity and freshness of its communication, such replay attacks cannot be prevented. Blocking updates of alias addresses also results in the reuse of these addresses. An attacker can then perform an active page scan for a particular device, and reuse an old alias address to successfully authenticate himself.

Wong and Stajano proposed a protocol to provide location privacy in Bluetooth networks [9]. It consists of three rounds and makes use of temporary pseudonyms. Each node in the network keeps a database of tuples containing his own temporary pseudonym, the pseudonyms of the other parties, and the shared link keys. If node A wants to communicate with node B, it selects a random nonce $R_1$, computes the hash $H_1$ using a hash function, and sends an $ID_1$ packet. The hash in the $ID_1$ packet hides the past pseudonym of node B. The latter can compute and verify the expected hash in the $ID_1$ packet using his database of the paired devices' temporary pseudonyms and their associated link keys with the nonce. When it successfully finds a match, it chooses a random nonce $R_2$, computes $H_2$, and responds with the $ID_2$ packet. On receiving the $ID_2$ packet, node A will verify the hash. If there is a match, node A will generate a random nonce $R_3$, compute the hash $H_3$ and reply with the $ID_3$ packet. On receipt of this message, node B will verify the hash $H_3$. After the protocol runs successfully, both parties update their temporary pseudonym. These new pseudonyms must be randomly generated. Wong and Stajano have suggested hashing some counter. The use of temporary pseudonyms helps to avoid location tracking. The security of the protocol depends on the randomness of the nonces, the irreversibility of the hash function and the secrecy of the shared link key. After the successful execution of the three-way protocol, both parties know they are communicating with the correct party. This protocol not provides full protection to location privacy attacks. An attacker can track easily stolen or lost devices.

In this paper, we propose to improve and to adapt the scheme proposed by Dave Singelée (figure 3) to provide the source and the sink location privacy in Wireless Body Area networks.

## 2. PROBLEM DEFINITION

### 2.1. Network model

We consider that the WBAN contains several sensor nodes that measure medical data such as ECG, body movement, temperature etc. (figure1 [5]). These sensor nodes have unique IDs. They have limited energy and memory space, and computation capability. These sensor nodes are also equipped with a radio interface and send their measurements wireless to a central device called the personal server or the base station or the sink.

Because the wireless body area network has a small size, we assume that all nodes of the network are in the range of the sink and can communicate directly with it. So, our network model has a star topology (figure 2).

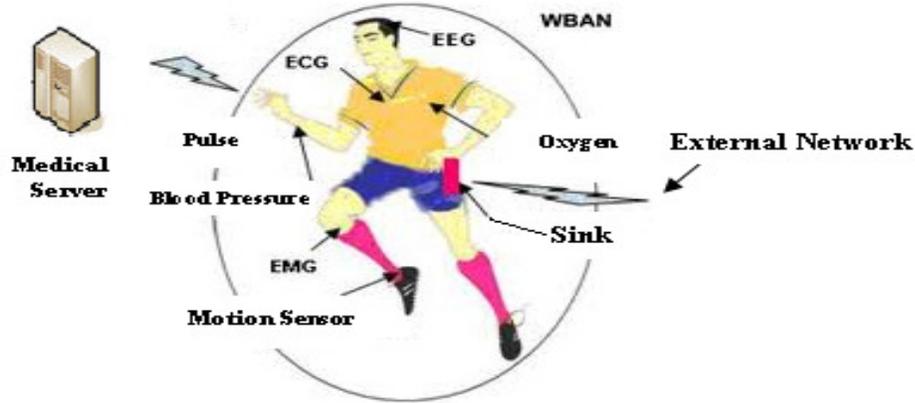

Fig.1. WBAN Architecture

The previous figure illustrates the general overview of the wireless body area network. There are several sensor nodes that collect medical data from the patient and send it to the sink. The sink is unique for each WBAN (and hence for every patient) and acts as a gateway between the WBAN and the external network. The external network can be any network providing a connection between the medical hub and the medical server. In most cases, the communication between the external network and the sink will be wireless. The medical server securely stores, processes and manages the huge amount of medical bio-data coming from all of the patients. This data can then be observed and analyzed by medical staff.

The following figure depicts our network model. All sensor nodes have the same level and can communicate directly with the sink. In the system there is also an attacker present who wants to track a particular user by the sensor nodes the latter is carrying.

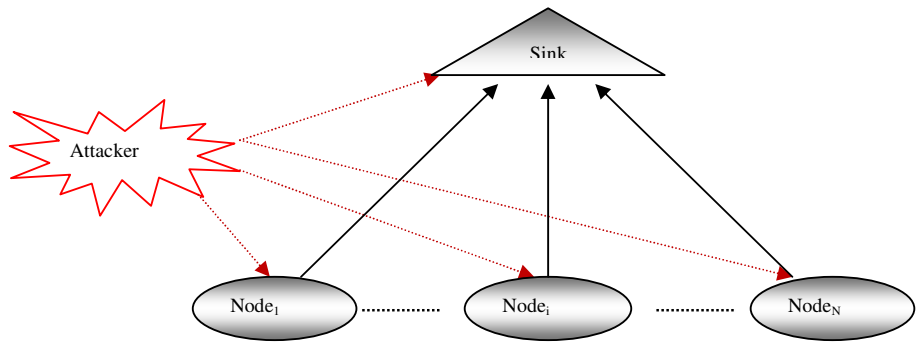

Fig. 2. Our Network Model

## 2.2 Security Assumptions

We assume that the sensor nodes are created with a Unique device Identifier (UId), which is known only by that particular sensor node. The UId of all the nodes has to be manually programmed into the base station and each UId acts as an initial shared secret between that device and the base station. The UId is used only during the bootstrapping process and is never exchanged in clear text, hence ensuring that this identifier is never explicitly disclosed to any other sensor node. Device tamper resistance mechanisms might have to be employed in order to ensure that the memory is flushed if any attempt is made to physically manipulate the device in order to retrieve this data.

## 2.3 Adversarial model

The model consists of the means of the adversary and his goals. The means of the attacker are represented using the following oracles [2]:
- Query Target or Query Sink: The attacker sends a message to the sink, and observes the response.
- Query node Ni: The attacker sends a message to the node Ni, and observes the response.
- Execute (Ni, Sink): The attacker forces Ni and the sink to communicate between them and eavesdrops on the exchanged messages.

During an attack game, the attacker is allowed to make a particular number of queries to each (or some) of the oracles. We parameterize the number of Query Sink messages by $q_s$, the number of Query node messages by $q_r$ and the number of Execute messages by $q_e$. An adversary with these means is denoted by $A[q_s, q_r, q_e]$ in the rest of the paper.

## 2.4 Attack games

The goal of an adversary in an attack game is twofold, the first is to distinguish between a node and the sink of the WBAN and the other is to detect which node/sink belongs to a specific WBAN.

To analyze the security of the protocol used to identify the source and the destination of messages, authors in [4] assume that its security level can be parameterized by a security parameter k and in the definition of parameterizable attack games, they used the notation poly(k) to represent any polynomial function of degree k.

### 2.4.1 Attack game 1

The goal of this attack game is to distinguish between a specific target T (the sink), chosen by the attacker, and another random node. The attack game goes as follows:

- The attacker selects a specific node $N_j = T$ from a particular WBAN. This will be the target node for the challenge.
- The attacker can query the three oracles (Query target T "Query Sink", Query node $N_i$, and Execute ($N_i$, T)). The numbers of allowed queries to these oracles are parameterized by $q_s$, $q_r$ and $q_e$ respectively.
- The adversary selects two nodes, $T_0$ and $T_1$. One of these nodes is equal to the target T (the sink), the other node is a random node $N_x$. The goal of the attacker is to indicate which one of these two nodes $T_b$ is the target node T (the sink).
- The attacker can query the three oracles (Query target $T_i$, Query node $N_i$, and Execute ($N_i$, T)).
- The attacker has to decide which node of $T_0$ and $T_1$ is equal to the target T (the sink).

An identification protocol P executed in a WBAN with security parameter k is $(q_s, q_r, q_e)$-location private if:

$\forall A[q_t, q_r, q_e]$ : Pr ($A[q_s, q_r, q_e]$ wins attack game 1 by guessing b) $\leq (1\backslash 2)+(1\backslash poly(k))$ [2, 6]

**Attack game 2**

The goal of this attack game is to detect that a certain node belongs to a specific WBAN. The attacker does not want to make a distinction between the nodes and the sink in the WBAN, detecting that a device (node/sink) is part of a specific WBAN is already enough. This attack makes sense from a practical point of view, since an attacker is typically not interested in detecting a specific device, but the user operating the device. And since a user is often carrying the same devices, which form the WBAN, this attack is sufficient to track the user.

The game goes as follows:

- The attacker selects a particular WBAN. This last is the target of the attacker.
- The attacker can query the two oracles Query node $N_i$ and Execute ($N_i$, T), as described previously. The numbers of allowed queries to these oracles are parameterized by $q_r$ and $q_e$ respectively.
- The adversary randomly selects one of the nodes $N_i$. This node is removed from the WBAN. The attacker also selects another node, which is not part of the same WBAN (and hence not known by the nodes $N_i$). These two nodes are randomly defined as T0 and T1. The goal of the attacker is to indicate which one of these two nodes $T_b$ belongs to the particular WBAN (and is hence known by the other nodes $N_i$).
- The attacker can query the three oracles (Query Sink, Query node $N_i$, and Execute ($N_i$, T)). The numbers of allowed queries to these oracles are parameterized by $q_s$, $q_r$ and $q_e$ respectively.
- The attacker has to decide which node $T_b$ (so $T_0$ or $T_1$) belongs to the WBAN formed by the nodes $N_i$ (the Sink is included). The attacker wins when his guess of the bit b was correct.

A protocol P executed in a WPAN with security parameter k is $(q_s, q_r, q_e)$-WBAN location private if:

$\forall A[q_s, q_r, q_e]$ : Pr ($A[q_s, q_r, q_e]$ wins attack game 2 by guessing b) $\leq (1\backslash 2)+(1\backslash poly(k))$ [2, 6]

Next is given our protocol design which aims to provide location privacy in wireless body area network.

## 3. DAVE SINGELÉE LOCATION PRIVACY PROTOCOL

This section presents Dave Singelée location privacy protocol in wireless personal area networks.

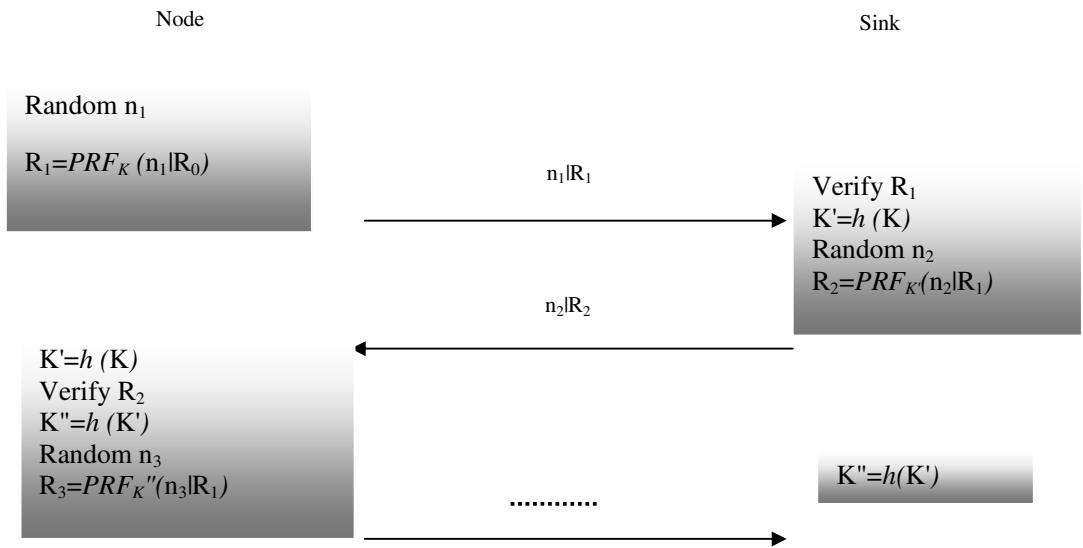

Fig. 3. Dave Singelée Location Privacy Protocol [2]

As depicted in the previous figure, author propose to compute the new temporary pseudonym from a random nonce and the old pseudonym using a pseudo random function "*PRF(.)*" and a shared key "K".

$R_{New} = PRF_K(n|R_{Old})$

After each round of the protocol, the key "K" is updated.
$K' = h(K)$

This scheme does not provide full protection against attack games because if a node and the sink use the same nonce, an adversary can win attack game 1 with a probability close to 100% by performing the following adversarial algorithm:

o The attacker selects a specific node N = T from a particular wireless personal area network. This will be the target node for the challenge.
o The attacker sends two queries to a node $N_i$, which shares an unknown key K with the target T. The current pseudonym shared by $N_i$ and T is R. In the first query, the node $N_i$ will reply with the pseudonym R. In the second query, the node $N_i$ will reply with the pseudonym $PRF_K(R|n)$.
o The adversary selects two nodes, $T_0$ and $T_1$. One of these nodes is equal to the target T, the other node is a random node $N_x$.
o The attacker sends a query to the nodes $T_0$ and $T_1$. This target query contains the pseudonym R.
o One of the nodes will reply to this query with a message containing the pseudonym $PRF_K(R|n)$, the other node with a random message. The node that has replied with $PRF_K(R|n)$ is the target node T.

Also, an adversary can win attack game 2 with a probability close to 100% by performing the following adversarial algorithm:

o The attacker selects a particular wireless personal area network, formed by the group of nodes $N_i$. This group is the target of the attacker.

o The adversary randomly selects one of the nodes $N_i$. This node is removed from the network. The attacker also selects another node, which is not part of this particular network (and hence not known by the nodes $N_i$). These two nodes are randomly defined as $T_0$ and $T_1$.

o The attacker sends two queries to both the nodes $T_0$ and $T_1$. One of the nodes will reply with the pseudonym R in the first query, and with the pseudonym $PRF_K(R|n)$ in the second query. The other node will reply twice with a random message (denoted by $X_1$ and $X_2$).

o The adversary randomly selects one of the nodes $T_b$ ($T_0$ or $T_1$), and sends the response of this node's first query (so R or $X_1$) in a query to each of the remaining (n − 1) nodes $N_i$ of the particular wireless personal network.

o If one of the nodes Ni replies with the pseudonym $PRF_K(R|n)$, the node $T_b$ is equal to the target node T. If all the nodes Ni send a random reply back not equal to $PRF_K(R|n)$, node $T_b$ is not part of the particular wireless personal area network and hence not the target node.

Next, is given our protocol design which aims to provide full protection against attack games. Our solution is designed to provide location privacy in wireless body area networks.

## 4. OUR PROTOCOL DESIGN

This section shows our location privacy protocol. First, is given the different notation used in our protocol; then is presented the detail of our protocol.

### 4.1 Notation

We will use the following notation to illustrate different Primitives in our protocol design:

- $n_1, n_2$… are examples of nonce.

- *Uid:* is the unique device identifier.

- *Temp:* is a template generated from *Uid*.

- *Id*t: is a temporary node's identifier.

- *h (m)*: a cryptographic hash function applied to the message *m.*
- *PRF(.)*: is a pseudo random function

- *M1|M2*: is the concatenation of messages *M1* and *M2*

### 4.2 Protocol description

In this subsection, is presented the different steps of our location privacy protocol for wireless body area networks.

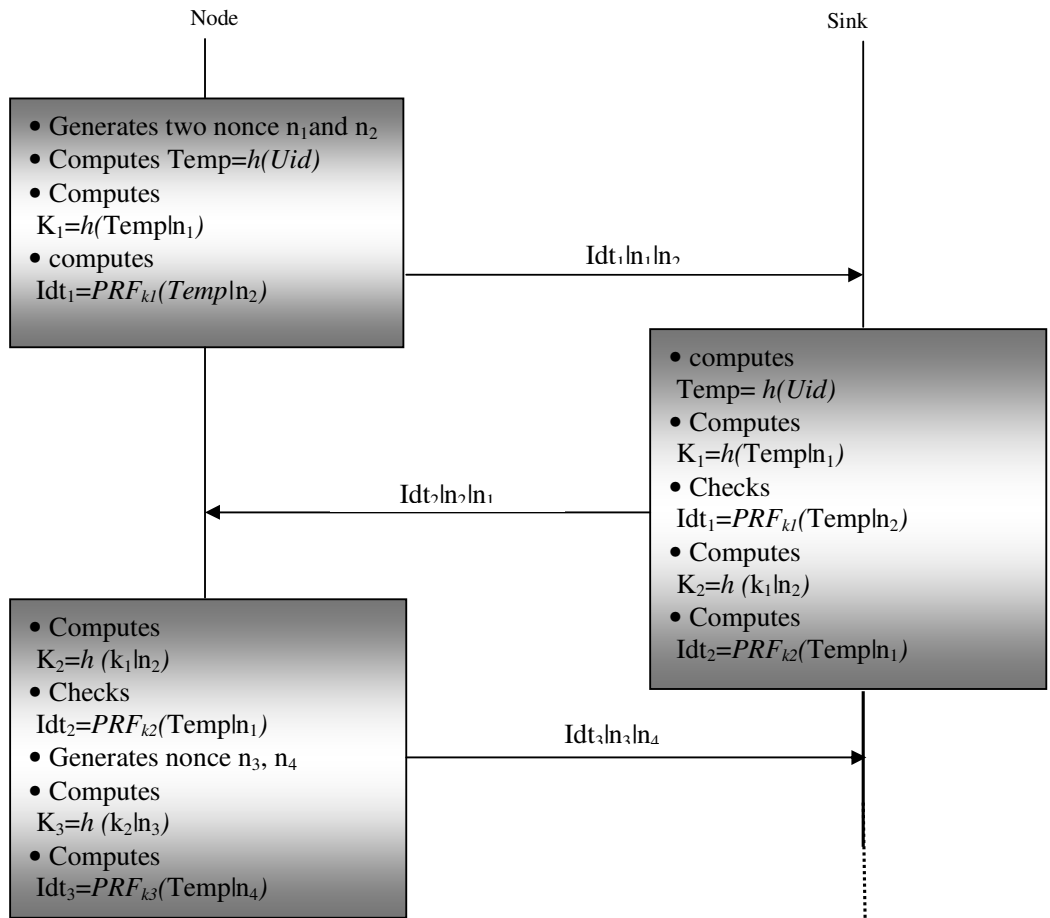

To provide location privacy in WBANs, nodes and the sink perform the following basic steps (as depicted in figure 4):

Fig. 4. Our Location Privacy Protocol

The node

**Step 1:** generates two nonces $n_1$ and $n_2$.

**Step 2:** generates a template "Temp" from Uid, Temp=$h\ (Uid)$.

**Step 3:** generates a session key from the template Temp and the nonce $n_1$ using the cryptographic hash function
$E\ (.)$, K= $h\ (Temp|n_1)$

**Step 4:** generates a temporary node's identifier Idt from $n_2$ and the template Temp using the key k and a pseudo random function $PRF\ (.)$.
Idt=$PRF_k\ (Temp|n_2)$

**Step 5:** transmits $n_1$, $n_2$ and Idt to the sink

The sink

---

**Step 6:** generates the template "Temp" from Uid,
Temp=$h\ (Uid)$.

**Step 7:** computes the session key K= $h\ (Temp|n_1)$

**Step 8:** checks Idt=$PRF_k\ (Temp|n_2)$

If the sink wants to send a message to the node, it computes the new key and the new pseudonym from the two received nonce using Temp and the hash function (show figure 4).
$K_2= h\ (K_1|n_2)$
$Idt_2= PRF_{K2}\ (Temp|n_1)$

---

After each transmission or reception, the node and the sink update the pairwise key "K" using the cryptographic hash function $"h\ (.)"$. The new key will be used to generate a new temporary node's identifier.

---

$K'= h\ (K|n)$ /* generation of the new key K' from the old
       key K using the cryptographic hash function
       $h\ (.)$*/

$Idt'= PRF_{K'}(Temp|n')$ /* generation of the new temporary
              node's identifier*/

---

### 4.3 Analysis of our location privacy protocol

First, we will examine and evaluate the efficiency of our location privacy protocol for WBANs against the two attack games presented in subsection 2.4. Then, we will analyze the energy needed for the execution of our proposed scheme.

#### 4.3.1 Efficiency against attack game1

To track a node or the sink, an adversary performs the following steps as described above (subsection 2.4):

○ The attacker selects a specific node $N_j = T$ from a particular WBAN. This will be the target node for the challenge.

○ The attacker sends two queries to a node $N_i$, which shares an unknown key K with the target T. The current pseudonym shared by $N_i$ and T is Idt. In the first query, the node $N_i$ will reply with the pseudonym Idt. In the second query, the node $N_i$ will reply with the pseudonym Idt'=$PRF_K'(Temp|n)$ where n is a random number and Temp is the template generated from the unique device identifier.

○ The adversary selects two nodes, $T_0$ and $T_1$. One of these nodes is equal to the target T, the other node is a random node $N_x$.

○ The attacker sends a query to the nodes $T_0$ and $T_1$. This target query contains the pseudonym Idt.

o One of the nodes (the target T) will reply to this query with a message containing the pseudonym Idt"=$PRF_{K''}$(Temp|n'), the other node with a random message X containing the pseudonym Idt'''.

Because the random responses of $N_i$, $T_0$ and $T_1$ which are respectively Idt', Idt" and Idt''', the adversary will not be able to detect which node is the target T.

The attacker is able to detect the target T if Idt"=Idt', but this will not be occur because the target and the nodes will never use the same nonces.

Our protocol is ($q_s$, $q_r$, $q_e$)-location private because:

$\forall A[q_t, q_r, q_e] : Pr(A[q_s, q_r, q_e])=0 \leq (1\backslash 2)+(1\backslash poly(k))$.

### 4.3.2 Efficiency against attack game2

To track a particular WBAN, an attacker performs the following steps as presented also in subsection 2.4.

o The attacker selects a particular WBAN, formed by the group of nodes $N_i$. This group (WBAN) is the target of the attacker.

o The adversary randomly selects one of the nodes $N_i$. This node is removed from the particular WBAN. The attacker also selects another node, which is not part of this particular WBAN (and hence not known by the nodes $N_i$). These two nodes are randomly defined as $T_0$ and $T_1$.

o The attacker sends two queries to both the nodes $T_0$ and $T_1$. One of the nodes will reply with the pseudonym Idt in the first query, and with the pseudonym Idt'=$PRF_K$(Temp|n') in the second query. The other node will reply twice with a random message (denoted by $X_1$ and $X_2$).

o The adversary randomly selects one of the nodes $T_b$ ($T_0$ or $T_1$), and sends the response of this node's first query (so Idt or $X_1$) in a query to each of the remaining (n −1) nodes $N_i$ of the particular WBAN.

The attacker wins attack game if one of the nodes $N_i$ replies with the pseudonym Idt' (the node $T_b$ is equal to the target node T), but this not will be occurred because all the nodes $N_i$ send a random reply back not equal to Idt'. The pseudonyms contained in the random replies are not equal to Idt' because the nodes do not use the same keys and the same nonces to compute their pseudonyms.

Our protocol is ($q_s$, $q_r$, $q_e$)-WBAN location private because:

$\forall A[q_t, q_r, q_e] : Pr(A[q_s, q_r, q_e])=0 \leq (1\backslash 2)+(1\backslash poly(k))$.

### 4.3.3 Energy consumption

Energy consumption is also taken into account. In our solution, we compute cryptographic hash values and use the result as an identifier (pseudonym). According to [7], the execution of cryptographic hash function requires 5,9µJ/Byte if the SHA-1 algorithm is used and the transmission and reception of a single byte of data requires 59, 2µJ and 28, 6µJ respectively.

Assuming that a 128-bit nonce and 128-bit Uid\Temp are used, the cost of computing the pseudonym "Idt" and updating the key "K" is 188,8 µJ.

The cost of transmitting or receiving two 128-bits identifiers (one of the sender and one of the destination) and two 128-bits nonce is 1894,4 µJ and 915,2 µJ respectively.

Therefore the total energy cost is 2998,4 µJ.

## 5. CONCLUDING REMARKS

Wireless Body Area Networks (WBANs) are an enabling technology for mobile health care. These systems reduce the enormous costs associated to patients in hospitals as monitoring can take place in real-time even at home and over a longer period. A critical factor in the acceptance of WBANs is the provision of appropriate security and privacy protection of the wireless communication medium. The data traveling between the sensors nodes should be kept confidential and integrity protected. Certainly in the mobile monitoring scenario, this is of uttermost importance.

In this paper, we have presented a light weight protocol to provide location privacy in wireless body area network. The basic idea of our solution consists on the use of temporary pseudonyms instead the use of hardware addresses to communicate in the wireless body area networks. This allows protecting the source and the destination of mobile devices in the WBANs.

Our protocol is efficient and energy saving.

### Acknowledgements

The research is developed in STIC Laboratory, Department of telecommunications, University of Tlemcen, Tlemcen, Algeria in collaboration and supervision of Professor Boucif Amar Bensaber, director of LAMIA Laboratory, Université du Québec à Trois-Rivières, Quebec, Canada.

This work was completed with the support of the natural sciences and engineering research council of Canada (nserc).